\documentclass[showpacs,onecolumn,superscriptaddress]{revtex4}

\usepackage{doi}%<----------
\usepackage{hyperref}
\hypersetup{
  colorlinks=true,        % false: boxed links; true: colored links
  linkcolor=blue,         % color of internal links
  citecolor=cyan,         % color of links to bibliography
}

\usepackage{graphicx}% Include figure files
\usepackage{dcolumn}% Align table columns on decimal point
\usepackage{bm}% bold math
\usepackage{color}

\usepackage{amsmath}
\usepackage{amssymb}

\begin{document}

\title{The role of dark matter halo in the evolution of the non-stationary disk of spiral galaxies}

\author{Karomat Mirtadjieva }
\email{karomat@astrin.uz}

\affiliation{Ulugh Beg Astronomical Institute, Astronomicheskaya 33, Tashkent 100052, Uzbekistan}
\affiliation{National University of Uzbekistan, Tashkent 100174, Uzbekistan}

\author{Kamola Mannapova}

\affiliation{National University of Uzbekistan, Tashkent 100174, Uzbekistan}

\date{\today}
\begin{abstract}
In this paper, we consider the problem of the evolution of the disk subsystem of
galaxies in view of the halo. To this end, we have studied the dependence of the
evolution of a non-linearly non-radially disk oscillating in its plane depending on the basic parameters of the dark matter halo numerically. The dark matter halo stabilizes the instabilities in the plane of the disk, but destabilizes its vertical oscillations. The global disk structure is dependent strongly on the mass and shape of the of dark matter halo. The evolutionary dependence of the oscillation process of a self-gravitating disk versus the indicated parameters of the dark matter halo are constructed.\end{abstract}
\maketitle

%\section{Introduction}

\section{Introduction}           %% first-level sections will be auto-capitalized
\label{sect:intro}

According to the results of the analysis of some observed data in theoretical researches of non-stationary evolution of disk-like subsystems of galaxies it is necessary to take into account the effect of halo (\citealt{Polyachenko+Fridman+1981, Malkov+1989, Zasov+etal+1990, Kuijken+1991}).In literature many works that are devoted to the construction of equilibrium disk-like models taking into account the halo, and also studying the stability of linear oscillations of the simplest models for revealing the role of halo are published (see, for example, (\citealt{Osipkov+Kutuzov+1987, Kutuzov+1991, Polyachenko+Shukhman+1979, Nuritdinov+1981}). It is definitively known that halo stabilizes the horizontal oscillations in a disk plane, but destabilizes its vertical oscillations. On the basis of this, Polyachenko and Shukhman (\citealt{Polyachenko+Shukhman+1979}) for the first time have revealed   a   narrow   interval   of admissible   values   of halo   mass   in spiral galaxies $(1.5–2.5)M_{d}\le M_{h}\le (3-4)M_{d}$, where $M_{d}$ is the mass of the disk. The lower bound of halo mass is defined from the condition of stability of an equilibrium disk relative to the bar-mode. According to our calculations have shown, in the case of a non-equilibrium disk the bar-mode is not the main cause of instability on obviously non-stationary stage of evolution (see, for example, (\citealt{Nuritdinov+etal+2008, Mirtadjieva+2012, Nuritdinov+1993} ). Therefore, we consider it necessary to study this problem for its non-linear oscillating models. With regret, it is necessary to notice that while modeling of non-equilibrium disks it is not possible to construct analytical nonlinear model to indicate the presence of a halo and to define the respective laws of non­stationary phenomena. This problem demands the numerical solving of the basic equations from the beginning, which complicates the performance of the model analysis. Let's notice that in reality, absolutely radial oscillations are a special case of non-radial fluctuations. Therefore, their study is possible only with the help of numerical methods.

\section{Basic relationships and equations}
\label{sect:Equations}

As we have noted above, till now the non-stationary phenomena in disk-like collision-less gravitating systems, taking into account a passive halo, have been studied close to their equilibrium states within the bounds of the theory of linear oscillations. Only in some papers (\citealt{Nuritdinov+1993}) for the first time the model of nonlinear non-radial osci11ating disk with surface density has been constructed:  

\begin{equation}
  \sigma \left( {x,y,t} \right) = {\sigma _0}{\left( {1 -\frac{{{x^2}}}{{{a^2}}} - \frac{{{y^2}}}{{{b^2}}}}\right)^{1/2}},
\label{eq:HaloI}
\end{equation}
and surrounded by passive ellipsoidal halo with gravitational potential

\begin{equation}
  {\Phi_h} =  - \frac{1}{2}{A_1}\left( e \right)\,\left( {{x^2} + {y^2}} \right) - \frac{{{A_2}\left( e \right)}}{2}{z^2}\,
\label{eq:HaloI}
\end{equation}
where $a$ and $b$ disk semi axes are time-dependent t; $A_1(e)$ and $A_2(e)$ are known functions of eccentricity e. The respective phase density of this model results in (\citealt{Nuritdinov+1993, Mirtadjieva+Nuritdinov+1997}) and looks like

\begin{equation}
  \Psi \left( {\vec r,\vec v,t} \right) = \frac{{{\sigma _0}}}{{2\pi {R_0}{{\left( {1 - {\Omega ^2}} \right)}^{1/2}}}}{f^{ - 1/2}}\chi \left( f \right)\
\label{eq:HaloI}
\end{equation}
where $\Omega$ is parameter of rotation, $\chi$ is Heaviside's  function and its argument is

\begin{equation}
  f = \left( {1 - {\Omega ^2}} \right)R_0^2 - S_1^{ - 1}{\left[ {N_2^{ - 1}\vec v - \left( {N_2^{ - 1}{N_1} + S_2^*} \right)\vec r} \right]^2} - \vec rK\vec r\
\label{eq:HaloI}
\end{equation}
moreover

\begin{equation}
  K = {N_2}\left[ {E + \left( {\Omega J - {N_3}} \right)S_1^{ - 1}\left( {\Omega J + {N_3}} \right)} \right]N_2^*,\
\label{eq:HaloI}
\end{equation}
where

\begin{equation}
  J = \left( {\begin{array}{*{20}{c}}0&1\\{ - 1}&0\end{array}} \right)\
\label{eq:HaloI}
\end{equation}
here $N_i$, $S_i$, and $K$ are two-dimensional matrices, from these matrices $N_1(t)$ and $N_3(t)$ are symmetric, the $*$ sign means transposition.

\begin{equation}
  S_1 = E + N_3^2 + \Omega \left( {J{N_3} - {N_3}J} \right),\quad \quad {S_2} = {N_2}\left( {\Omega J - {N_3}} \right)\
\label{eq:HaloI}
\end{equation}

In a perturbed state the elliptic disk has gravitational potential

\begin{equation}
  \Phi \left( {x,y,t} \right) =  - \vec r{\Phi _1}\vec r,\quad \quad {\Phi _1} = \left( {\begin{array}{*{20}{c}}{\rm A}&0\\0&{\rm B}\end{array}} \right)\
\label{eq:HaloI}
\end{equation}
where A and B are functions of $a(t)$ and $b(t)$.

The equations for matrices $N_i$ have been obtained in (\citealt{Nuritdinov+1993}), but they have not been numerically solved. In this work, we continue their analysis and corresponding calculations are mentioned below by the introduction of a parameter $p=M_{halo}/M_{disc}$ in the form of the ratio of masses of halo and disk. Then we have:

\begin{equation}
  \left\{ \begin{array}{l}\frac{{{\rm{d}}{{\rm{N}}_{\rm{1}}}}}{{{\rm{dt}}}}{\rm{ + N}}_{\rm{1}}^{\rm{2}}{\rm{ + 2}}{{\rm{K}}_{\rm{2}}}{\rm{ + p}}{{\rm{A}}_{\rm{1}}}{\rm{E = 0}}\\\frac{{{\rm{d}}{{\rm{N}}_{\rm{2}}}}}{{{\rm{dt}}}}{\rm{ + }}{{\rm{N}}_{\rm{1}}}{{\rm{N}}_{\rm{2}}}{\rm{ = 0}}{\rm{,}}\quad \\\frac{{{\rm{d}}{{\rm{N}}_{\rm{3}}}}}{{{\rm{dt}}}}{\rm{ + N}}_{\rm{2}}^{\rm{*}}{{\rm{N}}_{\rm{2}}}{\rm{ = 0}}\quad \quad \quad \end{array} \right.
\label{eq:HaloI}
\end{equation}
There is one inconvenient form in an initial state at t=O that the elements of matrices $N_1$, $N_2$, $N_3$ become discontinuous functions. That is why for carrying out of numerical integration we will transfer to the continuous functions of matrices $U$, $S$, $T$, $D$, which are defined by the following relations:

\begin{equation}
  U = N_2^{ - 1},\quad S = {N_3}N_2^{ - 1},\quad T = N_2^{ - 1}{N_1},\quad D = \frac{d}{{dt}}\left( {{N_3}N_2^{ - 1}} \right)\
\label{eq:HaloI}
\end{equation}
Then the evolution equations of a disk will look like follows:

\begin{equation}
  \begin{array}{l}\frac{{dU}}{{dt}} = T\;,\quad \\\frac{{dT}}{{dt}} =  - 2U{K_2} - p{A_1}U,\\\frac{{dS}}{{dt}} = D\;,\quad \\\frac{{dD}}{{dt}} =  - 2S{K_2} - p{A_1}S\end{array}\
\label{eq:HaloI}
\end{equation}
Thus, we have obtained a system of equations of evolution of a self-gravitating disk taking into account the halo in the form of matrix differential equations. Next, we numerically solve the Cauchy problem with given initial conditions.

\section{Analysis of the evolution equation for an unsteady disk with halo}

For integrate of this system represented by the matrix differential equations (11), obviously, it is necessary to specify physically acceptable initial conditions.
With this purpose we will introduce a matrix

\begin{equation}
  {\rm H} = \left( {\begin{array}{*{20}{c}}\mu &0\\0&{{\mu ^{ - 1}}}\end{array}} \right)\,
\label{eq:HaloI}
\end{equation}
where $\mu >1$, that characterizes the relation of a perturbed big semi axis $'a'$ to unperturbed one $'a_{o}'$ in initial moment of time $t_{0}=0$. Then the relation between the perturbed and unperturbed coordinates and velocities $r$, $v$ and $r_0$, $v_0$ it is possible to present in the form

\begin{equation}
  \vec r = H{\vec r_0} ~~~~~~~~ and  ~~~~~~~~ \vec v = {H^{ - 1}}{\vec v_0}\
\label{eq:HaloI}
\end{equation}

It turns out that a value of  $\mu$ characterizes the nonlinear deviations from the unperturbed equilibrium state, for which $\mu = 1$.

\begin{figure}
   \centering
  \includegraphics[width=14.5cm, angle=0]{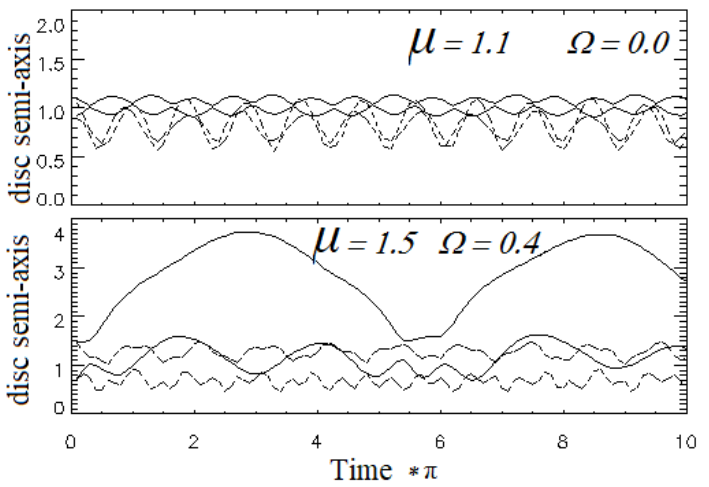}
  % \begin{minipage}[]{85mm}
   \caption{The behavior of the semi-axes of the disk, taking into account the halo for various values of the main parameters of the model. Here, the dashed lines denote the behavior of the semi-axes for the case $p = 1$, and continuous for $p = 0$.} 
%\end{minipage}
   \label{Fig1}
   \end{figure}

 Consequently, at the initial time

\begin{equation}
  \quad {\left( {{N_1}} \right)_0} \equiv 0,\quad \quad {\left( {{N_2}} \right)_0} \equiv {H^{ - 1}}\quad \quad {\left( {{N_3}} \right)_0} \equiv 0\
\label{eq:HaloI}
\end{equation}

We have developed a method for the numerical analysis of the equation of evolution of the disk of galaxies taking into account the halo (11). Numerical calculations were performed for various values of the parameter $p\in [0; 1]$, the initial perturbation $\mu\in [1; 2]$ and the circular rotation speed $\Omega\in [0;1]$ of the disk model. The calculations of the evolution of the disk model were carried out on the time interval $[0; 10\pi]$ with relative integrate accuracy $\Delta =10^{-7}$. According to numerical calculations, the dependencies of the major and minor semi-axes of the disk on the time $a(t)$ and $b(t)$ are found for various values of the parameters $p$, $\mu$ and $\Omega$ (Fig. 1). Values $a$ and $b$ were printed at intermediate points $t_{i}=\pi i/20$. The oscillation period was calculated by the counting method

\begin{equation}
  T = \frac{{{t_{\max \;n}} - {t_{\max \;1}}}}{n},\
\label{eq:HaloI}
\end{equation}
where $n$ is the number of maxima.
   
For each of the cases, a statistical amplitude $\varepsilon$ characterizing the degree of deformation of the system is also determined. The dependence of the statistical amplitude on the initial perturbation $\varepsilon (\mu)$ and the circular rotation speed $\varepsilon (\Omega)$ is obtained.

\section{Conclusion}
\label{sect:conclusion}
We list the main results we obtained in this paper:

1. A system of equations of evolution of a self-gravitating disk is obtained taking into account the halo in the form of matrix differential equations and a method for its numerical analysis is developed.

2. Numerical calculations were performed for various values of the initial perturbation $\mu$, the circular rotation speed $\Omega$ of the model, and the parameter $p$, which is the mass ratio of the halo and the disk. The corresponding critical values of the parameters are found.

3. The time dependence of the major and minor semi-axes of the disk are found on $a(t)$ and $b(t)$ for various values of the system parameters.

4. The statistical amplitude $\varepsilon$ characterizing the degree of deformation of the system is determined and the dependence of the statistical amplitude on the initial perturbation $\varepsilon (\mu)$ and the circular rotation speed $\varepsilon (\Omega)$ is obtained.

\bibliographystyle{apsrev4-1}  %% BibTeX style
\bibliography{bibtex}

\end{document}